\documentclass[final] {aipproc}
\usepackage{sidecap,graphicx,boxedminipage,epsfig,wrapfig,floatflt,shadow}
\layoutstyle{8x11double}

\begin{document}

\title{Cognitive Issues in Learning Advanced Physics: An Example from Quantum Mechanics}


\author{Chandralekha Singh and Guangtian Zhu}{
  address={Department of Physics and Astronomy, University of Pittsburgh, Pittsburgh, PA, 15260, USA}
}

\begin{abstract}
We are investigating cognitive issues in learning quantum mechanics in order to develop effective teaching and learning tools.
The analysis of cognitive issues is particularly important for bridging the gap between the quantitative and conceptual aspects of 
quantum mechanics and for ensuring that the learning tools help students build a robust knowledge structure.
We discuss the cognitive aspects of quantum mechanics that are similar or different from those of introductory physics and their implications
for developing strategies to help students develop a good grasp of quantum mechanics.

\end{abstract}

\maketitle

\section{Challenges in Classical vs. Quantum Mechanics}

The laws of physics are framed in precise mathematical language. 
Mastering physics involves learning to do abstract reasoning and making inferences using these abstract laws of physics
framed in mathematical forms. 
The answers to simple questions related to motion can be very sophisticated requiring a long chain of reasoning.
It is not surprising then that developing a solid grasp of physics even at the introductory level can be challenging. 

Learning quantum mechanics is even more challenging [1-12].
Unlike classical mechanics, we do not have direct experience
with the microscopic quantum world. Also, quantum mechanics has an abstract theoretical framework in which the most fundamental
equation, the Time-Dependent Schroedinger Equation (TDSE), describes the time evolution of the wave function or the state of a quantum system
according to the Hamiltonian of the system. 
This wave function is in general complex and does not directly represent a physical entity. However, the wave function at a given time can be 
exploited to make 
inferences about the probability of measuring different physical observables associated with the system. For example, the absolute square of the
wave function in position-space is the probability density.
Since the TDSE does not describe the evolution or motion of a physical entity, unlike Newton's second law, the modeling of the microscopic world in 
quantum mechanics is generally more abstract than the modeling of the macroscopic world in classical mechanics.

Quantum theory provides a coherent framework for reasoning about microscopic phenomena and has never failed to explain observations
if the Hamiltonian of the system is modeled appropriately to account for the essential interactions. 
However, the conceptual framework of quantum mechanics is often counter-intuitive to our everyday experiences.
For example, according to the quantum theory, the position, momentum, energy and other observables for a quantum mechanical entity
are in general not well-defined. We can
only predict the probability of measuring different values based upon the wave function when a measurement is performed.
This probabilistic interpretation of quantum mechanics, which even Einstein found disconcerting, is challenging for students.

Moreover, according to the Copenhagen interpretation of quantum mechanics, which is widely taught to students, the measurement of a physical
observable changes the wave function if the initial wave function is not an eigenfunction of the operator corresponding to the observable
measured. Thus, the usual time evolution of the system according to the TDSE is separated from what happens during the measurement of an observable.
Students often have difficulty with this notion of an instantaneous change or ``collapse" of the wave function during the measurement.
Our prior research~\cite{my6} shows that many students have common alternative conceptions about the collapse of the wave function during the measurement, 
e.g., many believe that the wave function gets stuck in the collapsed state after the measurement or it must go back to the original wave function if one 
waits long enough after the measurement. We found that when students were given the possibility that the wave function may neither stay stuck nor
go back to the original wave function, many students had difficulty understanding how anything other than those two outcomes was possible. It was clear 
from the discussions that the students had not internalized that after the measurement, the wave function will again evolve according to the TDSE 
starting from the collapsed wave function~\cite{my6}.

In quantum theory, position and momentum are not independent variables that evolve in a deterministic manner but are operators
in the Hilbert space in which the state of the system is a vector. 
For a given state of the system, the probabilities of measuring position or momentum in a narrow range depend on each other.
In particular, specifying the position-space wave function that can help us determine the probability of measuring the position in a narrow range 
specifies (via a Fourier transform) the momentum-space wave function that tells us the probability of measuring the momentum in a narrow range.
The eigenstates of the position or momentum operators span the Hilbert space so
that any state of the system can be written as a linear combination of a complete set of position eigenstates or momentum eigenstates.
The measurement of position (or momentum) yields a position (or momentum) eigenvalue with a certain probability depending upon the state of the system.
These concepts are challenging for students~\cite{my6}.

In addition to the lack of direct exposure to microscopic phenomena described by quantum theory and the counter-intuitive nature of the theory, 
the mathematical facility required in quantum mechanics can increase the cognitive load and make learning quantum mechanics even
more challenging. The framework of quantum mechanics is based on linear algebra. In addition, a good grasp of differential equations, special functions, 
complex variables etc. is highly desired. If students are not facile in mathematics, they may become overwhelmed by the mathematical details and
may not have the opportunity to focus on the conceptual framework of quantum mechanics and build a coherent knowledge structure.
Our earlier research~\cite{my6} shows that a lack of mathematical facility can hinder conceptual learning. Similarly,
alternative conceptions about conceptual
aspects of quantum mechanics can lead to students making mathematical errors that they would otherwise not make in a linear 
algebra course~\cite{my6}.

Many of the alternative conceptions in the classical world are over-generalizations of everyday experiences to
contexts where they are not applicable. For example, the conception that motion implies force often originates
from the fact that one must initially apply a force to an object at rest to get it moving. People naively over-generalize
such experiences to conclude that even an object moving at a constant velocity must have a net force acting on it.
One may argue that quantum mechanics may have an advantage here because the microscopic world does not directly deal with
observable phenomena in every day experience so students are unlikely to have alternative conceptions.
Unfortunately, that is not true and research shows that students have many alternative conceptions about quantum 
mechanics [1-12].
These conceptions
are often about the quantum mechanical model itself and about exploiting this model to infer what should happen in a given situation.
Students often over-generalize their intuitive notions from the classical world to the quantum world which can lead to incorrect inferences.

\begin{figure}
\epsfig{file=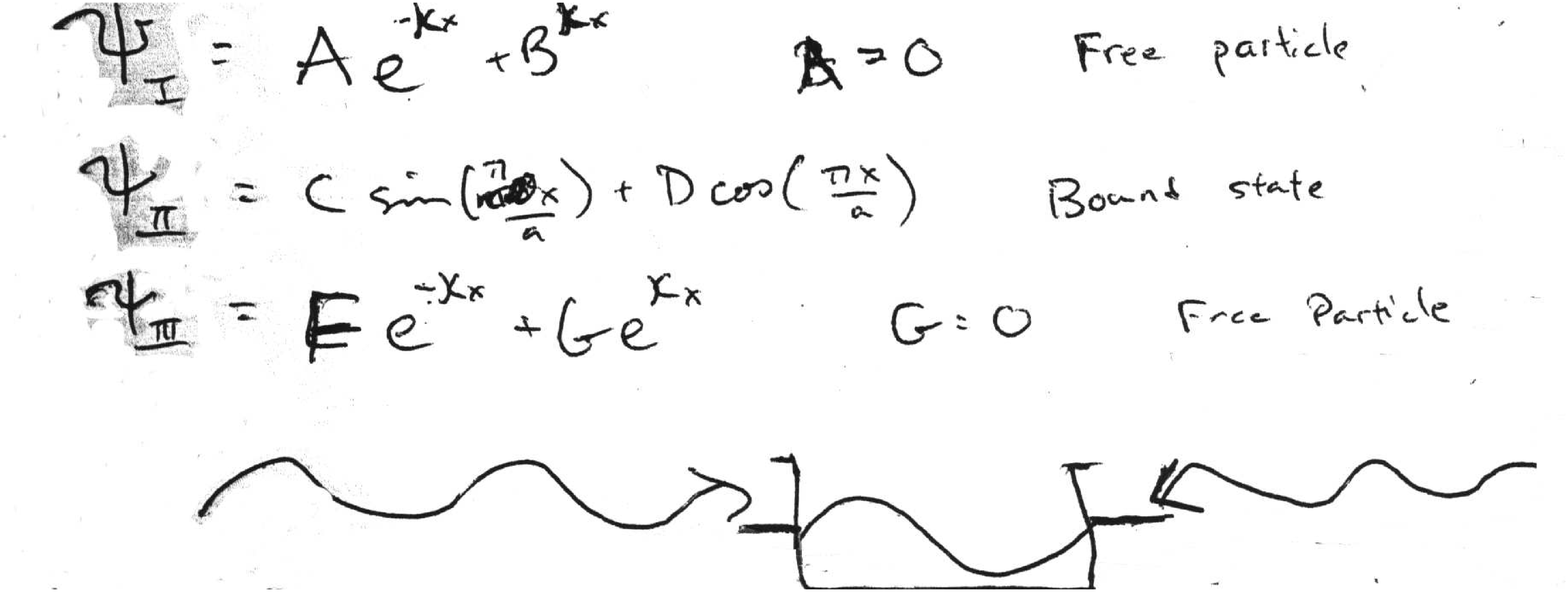,height=1.13in}
\caption{Bound and scattering states in the same plot}
\end{figure}

\begin{figure}
\epsfig{file=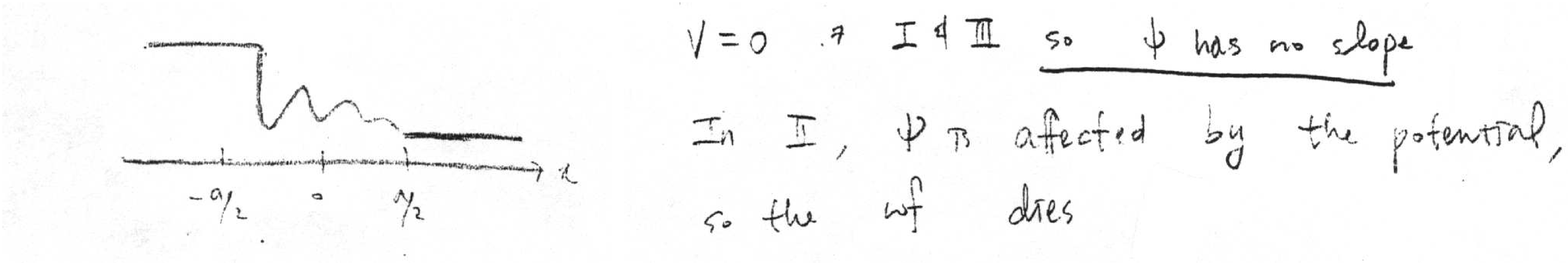,height=1.39in}
\caption{V=0 in regions I and III so the wave function has no slope and it is affected by the potential in region II so it dies}
\end{figure}

\begin{figure}
\epsfig{file=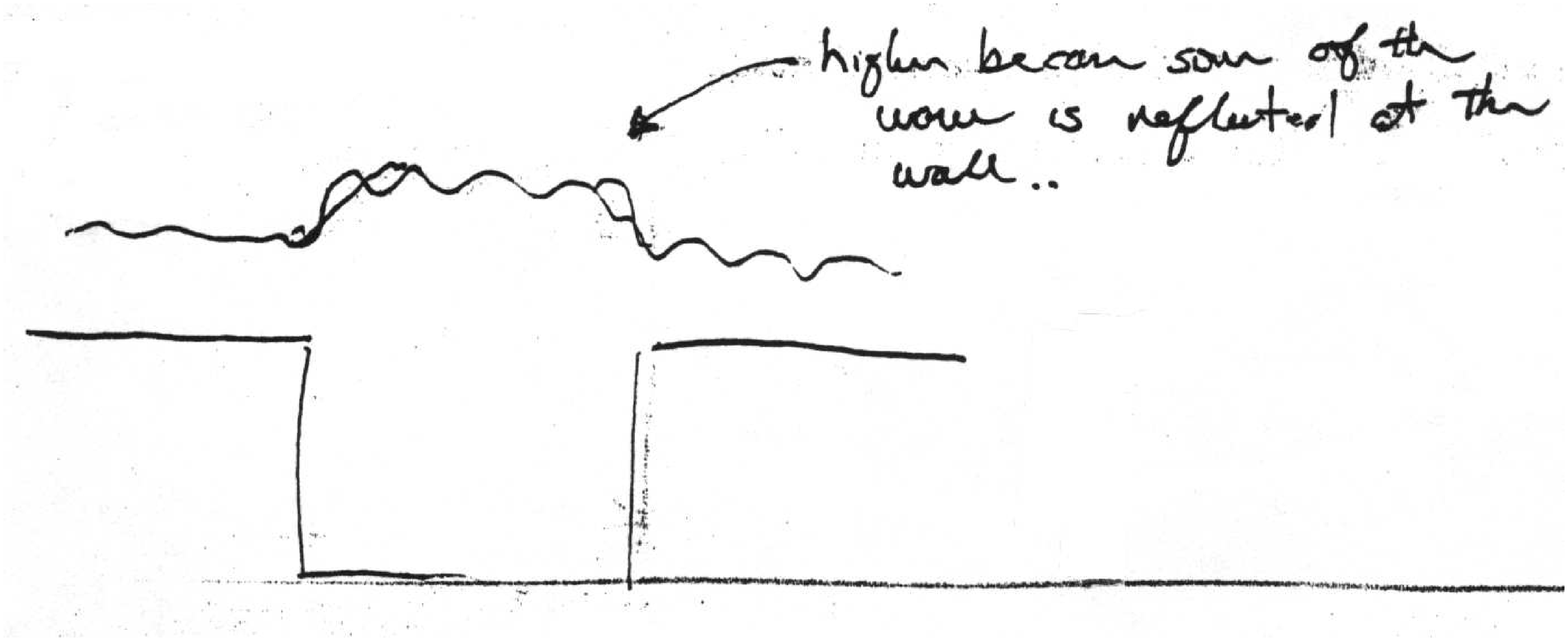,height=1.26in}
\caption{Higher in the well due to reflection}
\end{figure}

\begin{figure}
\epsfig{file=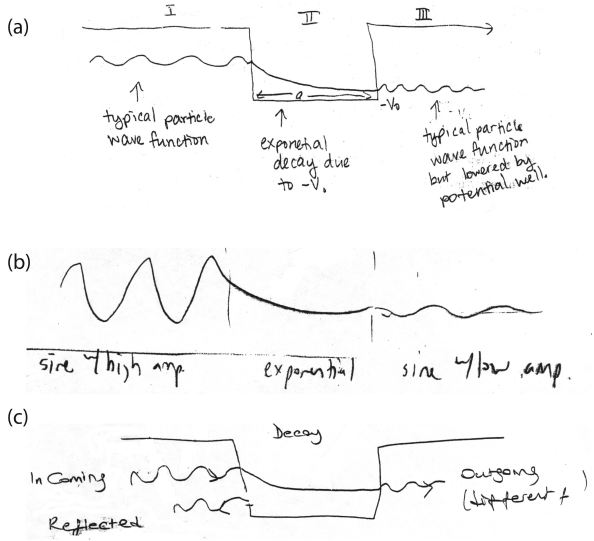,height=2.9in}
\caption{Exponential decay inside the well}
\end{figure}

\vspace{-0.15in}
\section{Investigation of Students' Difficulties with Wave Function}
\vspace{-0.07in}

\begin{figure}
\epsfig{file=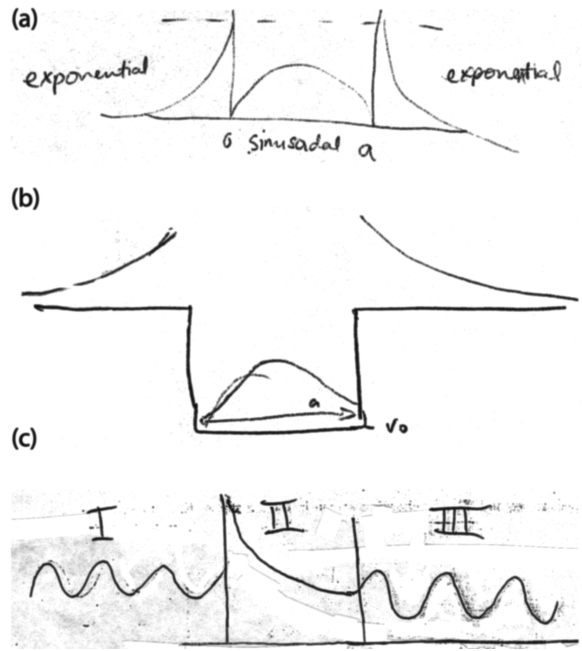,height=3.4in}
\caption{Discontinuity in the wave functions}
\end{figure}

\begin{figure}
\epsfig{file=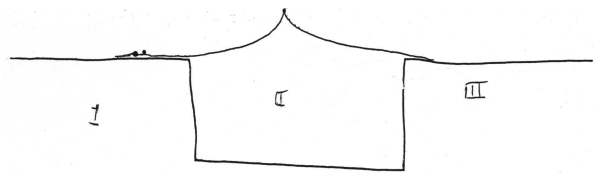,height=0.87in}
\caption{Cusp in the wave function}
\end{figure}

As discussed earlier, the wave function is central to quantum mechanics~\cite{jolly,sam}.
Here, we discuss an investigation of difficulties with the wave function that was carried out by administering written surveys to more than 
two hundred physics graduate students and advanced undergraduate students enrolled in quantum mechanics courses and
by conducting individual interviews with a subset of students.
Students were given a potential energy diagram for a one-dimensional finite square well of width $a$ and depth $-V_0$ between $0 \le x \le a$.
They were asked to draw a qualitative sketch of (a) the ground state wave function, (b) any one scattering state wave function and comment on the shape of 
the wave function in each case in all the three regions $x \le 0$, $0 \le x \le a$ and $x \ge a$.
The individual interviews were carried out using a think aloud protocol.
During the semi-structured interviews, students were asked to verbalize their thought processes while they answered
the questions. They were not interrupted unless they remained quiet for a while.
In the end, we asked them for clarifications of the issues they did not make clear. Here, we cite examples of students' difficulties.

We note that students were provided separate spaces for drawing the bound and scattering state wave functions so that
they do not confuse the vertical axis in the potential energy diagram given with the vertical axis of their sketch of the wave function.
But instead of simply showing the location of $x=0$ and $x=a$ in their sketches, 
many students redrew the potential energy diagram, situated their wave function in the potential energy well and did not specify what the
vertical axes of their plots were.

In response to the question, one interviewed student claimed that it is impossible to draw the bound and scattering state wave functions for a 
finite square well because one must find the solution of a transcedental equation which can only be solved numerically. When the student
was encouraged to make a qualitative sketch, he drew two coordinate axes and then drew some parallel curves and a straight line from
the origin intercepting the curves. He claimed that all he can say without solving the equation on the computer is that the intercepts will
give the wave function. While one must solve a transcendental equation to find the finite number of bound states for a finite square well, 
the student was asked to draw a qualitative sketch of the wave function, something that is taught even in a modern physics course.
In particular, students are taught that the bound state wave functions for a finite square well look sinusoidal inside the well with an exponential
tail outside in the classically forbidden region. It appeared that the student had memorized a procedure but had not developed a qualitative ``feel"
for what the bound and scattering state wave functions should look like for a finite square well.

Figure 1 shows a sketch from a student who incorrectly believed that the bound and scattering states can be part of the same wave function.
He felt that the sinusoidal wave function inside the well was the bound state and the part of the wave function outside the well was
the scattering state and corresponded to the "free particle". Some interviewed students claimed that the shapes of the various
bound state wave functions for the finite square
well cannot be sinusoidal inside the well since only the infinite square well allows sinusoidal bound states.
One student incorrectly claimed that the ground state of the finite square well should be Gaussian in shape to ensure that the wave function
has no cusp and exponentially decays to zero outside the well.

Figure 2 shows a sketch of the scattering state wave function by a student who incorrectly
claimed that the wave function has no slope because the potential is zero in regions I and III. 
While the probability density may be uniform, the wave function cannot be constant in those regions. 
The student also incorrectly believed that that the constant value of the wave function is lower in region III compared to region I since it
is affected by the potential in region II and dies. 
Figure 3 shows a sketch of the scattering state by a student who incorrectly
drew the wave function to be higher in region II and claimed: ``higher because some of the wave is reflected at the wall". Figure 4 shows
sketches by three students all of whom incorrectly believed that the wave function will decay exponentially in region II. These students have not learned
what one should observe when the potential energy is lower in the well in region II. Instead, they plotted a decaying wave
function from rote memory that may
correspond to a potential barrier. Moreover, similar to a student's sketch in Figure 3, the student who drew Figure 4(a) incorrectly claimed: ``typical
particle wave function but lowered by potential well" as though the oscillations in regions I and III should be around different references. 
These types of confusions are partly due to the inability to distinguish between the vertical
axis of the potential well (which has the units of energy) with the vertical axis when drawing the wave function. Also, in Figure
4(c), the student drew the incoming and reflected waves separately in region I but only drew the incoming part to be continuous with the wave function in region
II which is incorrect. 
Figure 5 shows three students' plots in which the wave functions drawn have discontinuities and Figure 6 shows a plot in which there is a cusp. 

Interviews and written explanations suggest that many students drew diagrams of the wave function from memory without thinking about the physical 
meaning of the wave function. This may partly be due to the fact that the wave function itself is not physical and cannot be observed experimentally.
Additional cognitive resources are required to make sense of the wave function in order to draw it correctly.
For example, a discontinuity in the wave function is not physical because the absolute square of the wave function is related to the 
probability density for finding the particle and a discontinuity at a point would imply that the probability of finding the particle will 
depend on whether we approach that point from the left or the right side. Similarly, the wave function cannot have a cusp because it would imply 
that the expectation value of the kinetic energy (related to the second derivative of the wave function) is infinite.

%\vspace*{-.2in}
\vspace*{-.1in}
\section{Conclusion and Outlook}
\vspace*{-.10in}

While quantum mechanics may require reasoning at the formal operational level in the Piagetian hierarchy of cognitive levels~\cite{piaget}, 
it is possible to design instruction that helps students develop intuition.
The notion of the ``zone of proximal development"~\cite{vygotsky} (ZPD) attributed to Vygotsky focuses on what a student can do on his/her own vs. 
with the help of an instructional strategy that accounts for his/her prior knowledge and skills and builds on it. In quantum mechanics,
we can exploit students' prior knowledge of probability and mathematical skills. 
But the non-intuitive nature of quantum mechanics and other issues discussed earlier imply that
scaffolding, which is at the heart of ZPD, is critical for helping students learn concepts.
Scaffolding can be used to stretch students' learning far beyond their initial knowledge by carefully crafted instruction.
We are taking into account these issues and students' prior knowledge to develop Quantum Interactive Learning Tutorials (QuILTs) and tools 
for peer-instruction~\cite{my3,my7}. 
These learning tools employ computer-based visualization tools and help students take advantage of the visual
representation of the quantum mechanical concepts, e.g., wave function, in order to develop intuition about quantum phenomena.  

\vspace*{-.20in}
\begin{theacknowledgments}
We thank NSF for PHY-0653129 and PHY-055434.
\end{theacknowledgments}
\vspace*{-.10in}

\vspace*{-.060in}
\bibliographystyle{aipproc}

\end{document}